\documentclass[10pt]{article}

\usepackage{amsmath}
\usepackage{amssymb}
\usepackage{graphicx}
\usepackage{cite}
\usepackage{color} 
\usepackage{gensymb}

\usepackage{lineno}

\usepackage{microtype}
\DisableLigatures[f]{encoding = *, family = * }

\usepackage[labelfont=bf,labelsep=period,justification=raggedright]{caption}

\makeatletter
\renewcommand{\@biblabel}[1]{\quad#1.}
\makeatother

\date{}

\newcommand{\micron}{\ensuremath{\mu\mathrm{m}}}

\newcommand{\e}{\mathbf{e}}
\renewcommand{\r}{\mathbf{r}}
\renewcommand{\v}{\mathbf{v}}
\newcommand{\B}{\mathrm{B}}
\newcommand{\C}{\mathrm{C}}
\newcommand{\D}{\mathrm{D}}
\newcommand{\G}{\mathrm{G}}
\renewcommand{\S}{\mathrm{S}}
\newcommand{\U}{\mathrm{U}}
\newcommand{\V}{\mathrm{V}}

\newcommand{\ol}[1]{\overline{#1}}

\begin{document}

\begin{flushleft}{
\Large\textbf{Shape mode analysis exposes movement patterns in biology:
flagella and flatworms as case studies\\
(accepted for publication in PLoS One)
}
}
\\
Steffen Werner$^{1}$, 
Jochen C. Rink$^{2}$, 
Ingmar H. Riedel-Kruse$^{3}$,
Benjamin M. Friedrich$^{1,\ast}$
\\
\bf{1} 
Max Planck Institute for the Physics of Complex Systems, Dresden, Germany
\\
\bf{2} 
Max Planck Institute for Cell Biology and Genetics, Dresden, Germany
\\
\bf{3}
Department of Bioengineering, Stanford University, Stanford, CA, USA
\\
$\ast$ E-mail: benjamin.friedrich@pks.mpg.de
\end{flushleft}

\section*{Abstract}
We illustrate shape mode analysis as a simple, yet powerful technique to
concisely describe complex biological shapes and their dynamics.
We characterize undulatory bending waves of beating flagella
and reconstruct a limit cycle of flagellar oscillations,
paying particular attention to the periodicity of angular data.
As a second example, we analyze non-convex boundary outlines of gliding flatworms,
which allows us to expose stereotypic body postures that can be related to two different locomotion mechanisms.
Further, shape mode analysis based on principal component analysis
allows to discriminate different flatworm species, 
despite large motion-associated shape variability. 
Thus, complex shape dynamics is characterized by a small number of shape scores that change in time.
We present this method using descriptive examples, explaining abstract mathematics in a graphic way.

\section*{Introduction}

Life presents itself in manifold morphologies. 
Quantifying morphology is often the first step to relate form and function. 
A common task in shape characterization amounts to finding those 
morphological features and geometric quantities with maximal descriptive power. 
This is especially challenging when aiming to understand shape changes 
of soft or flexible structures, such as beating cilia or animals without rigid skeletons.
Shape mode analysis is a standardized way to find such 
quantities \textit{a posterori}, after data collection, 
by combining a large number of partially redundant morphometric features into 
a small set of distinct shape scores
\cite{Pearson:1901,Abdi:2010,Stephens:2010,Jolliffe:2005,Jackson:2005}. 

Shape mode analysis is a well-known technique in engineering and computer
science, \textit{e.g.} for image recognition \cite{Cootes:1995}, yet only recently
researchers began to apply it to biological data sets. 
One of the earliest application of this method to biological shape data 
was by Sanger \textit{et al.}, analyzing human arm posture \cite{Sanger:2000}.
Pioneered by Ryu \textit{et al.}, shape mode analysis has been particularly used 
to analyze motility patterns of the round worm \textit{C. elegans} 
\cite{Stephens:2008,Stephens:2010,Raviv:2010,Gallagher:2013}. 

Here, we adapt principal component analysis to analyze 
and quantify movement patterns in two 2D image data sets:
(i) the bend centerline of beating flagella, 
and (ii) the closed boundary outline of gliding flatworms. 
We reconstruct a limit cycle of flagellar oscillations
using a data set from swimming bull sperm,
which allows us to study not only regular flagellar oscillations,
but also noisy deviations from perfect periodicity,
thereby contributing to the characterization of the flagellum as a noisy oscillator \cite{Ma:2014,Wan:2014arXiv}. 

In contrast to flagella or the slender shapes of the round worm \textit{C. elegans}, 
many cells and organisms display morphologies that are more suitably described by their outline contour. 
However, outline contours can vary dramatically in the absence of skeletal elements, as is the case in planarian flatworms. 
Planarians have recently become an important model system for regeneration and growth dynamics \cite{Newmark:2002}. 
Their flattened and elongated body plan morphology is kept in shape 
by a deformable extracellular matrix material and the contraction status of their muscular plexus. 
Many species exist worldwide that often differ in body shape. 
However, measuring body shape in behaving animals is challenging, 
because changes in muscle tone constantly change the projected body shape and still images 
therefore rarely capture the ``true'' shape of the animal. 
Accurate quantification of shape in fixed specimens is similarly problematic, 
owing to various contraction artifacts of the fixation methods. 
We therefore thought to explore shape mode analysis 
with respect to its utility in extracting average shape information from movie sequences of living animals. 
As a first test, we analyzed an extensive high-precision tracking data set of gliding flatworms.
Planarians display a smooth gliding motility, 
resulting from the coordinated beat patterns of the cilia in their densely ciliated ventral epithelium \cite{Rink:2009, Rompolas:2010}. 
We find that a bending mode correlates with active turning during gliding motility,
showing that steering is achieved by a bending of the long body axis.
Additional modes characterize stereotypic width changes of these worms not reported before.
These width changes are shown to become particularly pronounced during a second type of motility behavior, inch-worming, 
normally associated to escape responses, but also observed in phenotypes with impaired cilia functionality \cite{Rink:2009, Rompolas:2010}. 
Our method reveals regular lateral contraction waves with a period of about $4\,\mathrm{s}$ in inch-worming worms.
We find that the extraction of body postures from tracked outline contours 
enables accurate shape measurements of flatworms, 
which we demonstrate by the ability to differentiate between different flatworm species. 
Supporting the notoriously difficult taxonomy of these soft-bodied animals 
with statistical quantification of genus- or species specific body shapes represents an interesting application of our method. 

By analyzing two typical classes of biological data sets in a pedagogical setting and
by explaining the mathematics in a graphic way,
we hope to provide an accessible account of this versatile method. 
Here, shape mode analysis is based on the mathematical technique of principal component analysis and
allows to project a multi-feature data set on a small set of 
empirical shape modes, which are directly inferred from the data itself. 
Principal component analysis thus represents a dimensionality-reduction technique,
where a big data set residing in a high-dimensional `feature space',
is projected from onto a convenient `shape space' of lower dimension with minimal information loss \cite{Jackson:2005,Jolliffe:2005}. 
As a side-effect, this method reduces measurement noise by averaging over several, partially redundant features.
The wide applicability of principal component analysis comes at the price of a 
diverse terminology across different disciplines, see Table~\ref{tab:terminology}.

For simplicity, we focus on linear principal component analysis in the main text. 
In an appendix, we discuss non-linear generalizations such as kernel methods 
and show how these can be used to analyze angular data, 
using the sperm data set as a descriptive example. 
Our analysis demonstrates how to relate organism shape and motility patterns 
in a pedagogical setting using flagella and flatworms as prototypical examples.

\section*{Results and Discussion}

\subsection*{A minimal example}

First, we discuss a minimal example to illustrate the 
key concept of dimensionality reduction by principal component analysis,
which forms the basis of our shape mode analysis approach.

\begin{figure}[!ht]
\begin{center}
\includegraphics[width=6in]{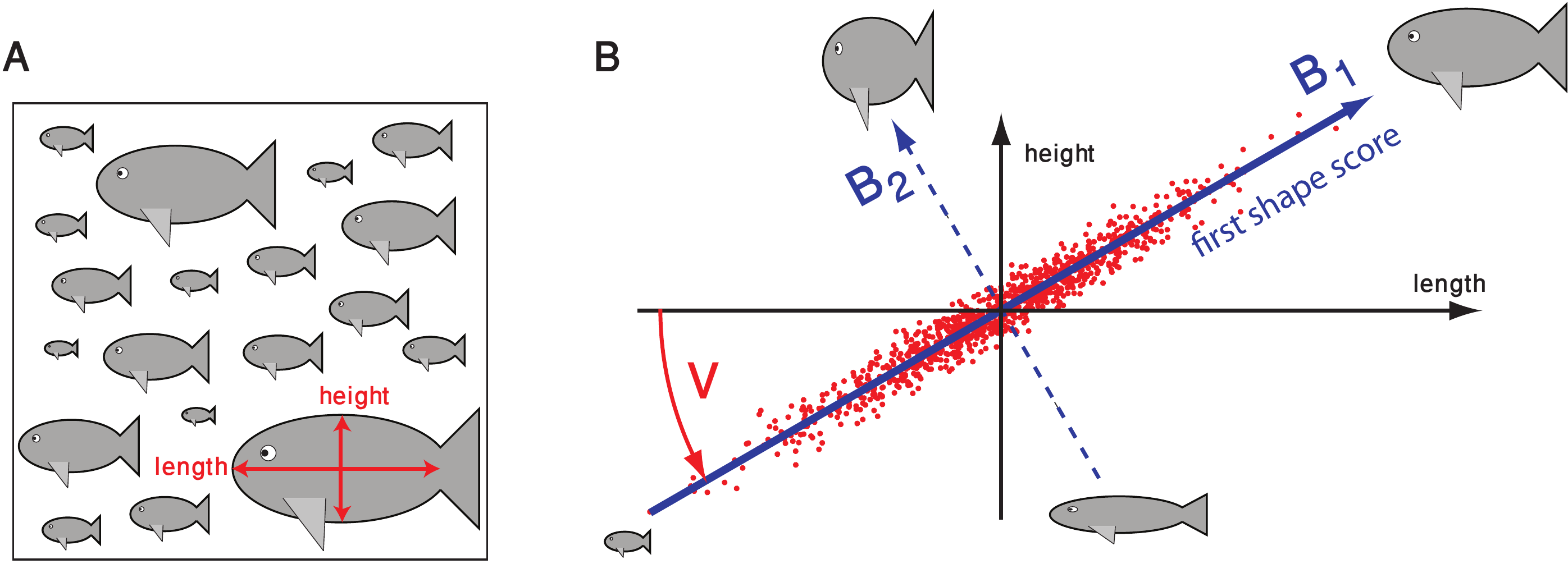}
\end{center}
\caption[]{
{\bf Illustration of principal component analysis.} 
\textbf{A.}
As a minimal example, 
we consider a hypothetical data set of 
length and height measurements for a collection of $n$ individuals,
\textit{i.e.} there are just $m=2$ geometric features measured here.
\textbf{B.} 
In this example, length and height are assumed to be strongly correlated,
thus mimicking the partial redundancy of geometrical features commonly observed in real data.
Principal component analysis 
now defines a change of coordinate system 
from the original (length,height)-axes (shown in a black) 
to a new set of axes (blue) that represent the principal axes of the feature-feature covariance matrix of the data.
Briefly, the first new axis $\v_1$ points in the direction of maximal data variability, 
while the second new axis $\v_2$ points in the direction of minimal data variability.
The change of coordinate system is indicated by a rotation $V=[\v_1,\v_2]$ around the center of the point cloud representing the data.
By projecting the data on those axes that correspond to maximal feature-feature covariance,
in this example the first axis, 
one can reduce the dimensionality of the data space, while retaining most of the variability of the data.
In the context of morphology analysis, we will refer to these new axes as `shape modes' $\v_i$, which 
represent specific combinations of features.
The new coordinates are referred to as `shape scores' $B_i$. 
}
\label{fig_rot}
\end{figure}

Assume we are given a data set
that comprises $m$ geometrical features measured for each of $n$ individuals,
say the distribution of length and height in a shoal of $n$ fish
such that $m=2$, see Fig.~\ref{fig_rot}A.
To mimic the partial redundancy of geometrical features commonly observed in real data, 
we further assume that these two features are strongly correlated, see Fig.~\ref{fig_rot}B. 

Principal component analysis now defines a unique change of coordinate system 
such that the new axes (blue) point along the principal directions of feature-feature covariance:
in the new coordinate system, the shape coordinates become linearly uncorrelated.
In the context of shape mode analysis, 
the new axes are called `shape modes' $\v_{i}$, 
while the corresponding coordinates will be referred to as `shape scores' $B_i$.
The first shape mode $\v_1$ points into the direction of maximal variation in the data.
In this example, the shape score $B_1$ corresponding to this first shape mode 
provides a robust measure of size that combines length and height measurements.
The remaining second shape mode $\v_{2}$ points along the direction of least covariance.
The corresponding shape score $B_2$ can be interpreted as an aspect ratio in this example.
As this second shape score $B_2$ displays only little variation, 
the data set is well described by just the first shape score $B_1$, 
which represents an effective dimensionality reduction from $m=2$ to one dimension.

We emphasize that the concept of an 
$n\times m$ measurement-feature matrix is rather generic and 
is encountered in many other contexts, such as measurements of
dynamic flagellar centerline shapes or flatworm outlines as discussed next.

\subsection*{Characterizing the flagellar beat as a biological oscillator}

Sperm cells are propelled in a liquid by regular bending waves of their 
flagellum, a slender cell appendage of $30-100\,\micron$ length 
\cite{Alberts:cell}. 
The flagellar beat is powered by ten-thousands of dynein molecular motors inside the 
flagellum that constantly convert chemical energy into mechanical work 
\cite{Howard:2001}. 
The regular shape changes of the flagellum determine speed and direction of 
sperm swimming \cite{Gray:1955b,Friedrich:2010}. 
Eukaryotic flagella propel also many other
microswimmers including green algae and ciliated Protozoans, or participate in fluid transport inside multicellular animals \cite{Sanderson:1981}.
Here, we analyze a data set of flagellar swimming of bull sperm 
\cite{Riedel:2007} using shape mode analysis.
Methods are described in \cite{Riedel:2007};
the frame-rate was $250\,\mathrm{Hz}$. 

\textit{Tangent angles characterize flagellar waves.}
In these experiments, sperm cells swam parallel to a boundary surface with an 
approximately planar flagellar beat. This effective confinement to two space 
dimension greatly facilitates tracking of flagellar shapes and their analysis. 
The (projected) shape of a bent flagellum at a time $t$ is described by 
the position vector $\r(s,t)$ of points along the centerline of the flagellum
for $0\le s\le L$, 
where $s$ is the arclength along the flagellar centerline
and $L\approx 58.3\micron$ the total flagellar length, see Fig.~\ref{fig_sperm}A.
To characterize shapes, we need a description that is independent of the 
actual position and orientation of the cell in space. To this end, we 
introduce a material frame of the sperm head consisting of the head center 
position $\r(t)$ and a unit vector $\e_1(t)$ pointing along the long axis of 
the prolate sperm head. 
The length of this long axis is $2r_1\approx 10\,\mu m$,
such that $\r-r_1\e_1$ corresponds to the proximal tip of the flagellum.
Additionally, we introduce a second unit vector $\e_2(t)$, which is 
obtained by rotating $\e_1$(t) in the plane of swimming by an angle of $\pi/2$ 
in a counter-clockwise fashion. With respect to this material frame, the 
tracked flagellar shape is characterized by a tangent angle $\psi(s,t)$ as \cite{Riedel:2007,Friedrich:2010}
\begin{equation}
\label{eq_psi}
\r(s,t)=\r(t) - r_1\e_1-\int_0^s ds'\Big( \cos\psi(s',t)\e_1(t) + \sin\psi(s',t)\e_2(t)\Big).
\end{equation}
This tangent angle measures the angle between the vector $\e_1$
and the local tangent of the flagellum at position $\r(s,t)$. 
Importantly, this tangent angle representation characterizes 
flagellar shape independent of cell position and orientation. 
Tracking a high-speed recording 
with $n=1024$ frames corresponding to time-points $t_1,\ldots,t_n$ and
using $m=41$ control points $s_j=j\,L/m$ along the flagellum, 
we obtain an $n\times m$ measurement matrix $\Psi$ for the tangent angle with 
$\Psi_{ij}=\psi(s_j,t_i)$. 
This matrix $\Psi$ represents a kymograph 
of the flagellar beat; an example is shown in Fig.~\ref{fig_sperm}B. 
The apparent stripe pattern reflects the periodicity of the flagellar beat. 
The slope of the stripes is directly related to the wave velocity of 
traveling bending waves that pass down the flagellum
from its proximal to its distal end.
On a more abstract level, the matrix $\Psi$ comprises $n=1024$ 
independent measurements (time points) of $m=41$ geometric features 
(tangent angle at the $m$ control points).

\textit{PCA decomposition of flagellar bending waves.}
We will now show how a set of principal shape modes can be extracted from this representation. 
First, we define a mean shape of the flagellum by averaging each column of 
the matrix $\Psi$, \textit{i.e.} we average over the $n$ measurements \cite{Riedel:2007}. 
The resultant mean tangent angle $\psi_0(s)$ and corresponding flagellar shape is shown 
in Fig.~\ref{fig_sperm}B. 
We note that taking a linear mean of angular data is admissible here,
since angles stay in a bounded interval and do not jump by $2\pi$;
a general procedure that can cope also with jumps of $2\pi$ is discussed in the appendix.
The mean tangent angle $\psi_0(s)$ is non-zero, which relates to an intrinsic asymmetry of the flagellar bending waves. 
Asymmetric flagellar beating implies swimming along curved paths \cite{Brokaw:1979,Friedrich:2010}. 
Cellular signaling can change this flagellar asymmetry \cite{Wood:2005} and has been assigned 
a crucial role in non-mammalian sperm chemotaxis \cite{Alvarez:2014}.
Here, we are interested in flagellar shape changes, \textit{i.e.} deviations from 
the mean shape. 
Thus, we devise an $n\times m$-matrix $\Psi_0=[\boldsymbol{\psi}_0;\ldots;\boldsymbol{\psi}_0]$ all of 
which rows are equal to the mean tangent angle $\boldsymbol{\psi}_0=[\psi_0(s_1),\ldots,\psi_0(s_m)]$. 
We can now compute the $m\times m$ feature-feature covariance matrix as
\begin{equation}
\label{eq_C}
\C=(\Psi-\Psi_0)^T(\Psi-\Psi_0),
\end{equation}
see Fig.~\ref{fig_sperm}C.
We find strong positive correlation along the main diagonal of this 
covariance matrix (dashed line), which implies that tangent angle 
measurements at nearby control points are correlated. 
This short-range correlation relates to the bending stiffness of the 
flagellum. It implies partial redundancy among the measurements corresponding 
to nearby control points along the flagellum. More interestingly, we find 
negative correlation between the respective tangent angles that are an 
arclength distance $\lambda/2$ apart. What does this mean? The flagellar beat 
can be approximated as a traveling bending wave with a certain wave length 
$\lambda$. This wavelength manifests itself as a ``long-range correlation'' in 
the covariance matrix $\C$.

We will now employ an eigenvalue decomposition of the $m\times m$ covariance matrix $\C$,
yielding eigenvalues $d_1,\ldots,d_n$ and eigenvectors $\v_1,\ldots,\v_n$ such that
$\C\v_i=d_i\v_i$.
In analogy to the minimal example above, we refer
to the eigenvectors $\v_i$ as shape modes. 
The shape modes $\v_i$ correspond to axes of a new coordinate system of feature space;
in this coordinate system, 
the variations of the data along each axis are linearly uncorrelated 
and have respective variance $d_i$ for axis $\v_i$, $i=1,\ldots,n$.
We can assume without loss of generality 
that the eigenvalues $d_1\ge\ldots\ge d_n$ of $\C$ are sorted in descending order, 
see Fig.~\ref{fig_sperm}D.
For the sperm data, we observe that the eigenvalue spectrum sharply drops after $d_2$; 
in fact, the first two shape modes $\v_1$ and $\v_2$ together account for 
$95\%$ of the variance of the data.
We now choose to deliberately chop the eigenvalue spectrum after $d_2$ and 
project the data set on the reduced ``shape space'' 
spanned by the shape modes $\v_1$ and $\v_2$. 
Generally, the mode-number cutoff will be application specific and requires supervision.
Formal criteria to chose the optimal cutoff have been discussed in the
literature, see \text{e.g.} \cite{Abdi:2010} and references therein.

Each recorded flagellar shape, 
that is each row $\Psi_i$ of the data matrix $\Psi=[\boldsymbol{\psi}_1;\ldots;\boldsymbol{\psi}_n]$ 
can now be uniquely expressed as a linear combination of the shape modes $\v_k$
\begin{equation}
\label{eq:decomp}
\boldsymbol{\psi}_i = 
\boldsymbol{\psi}_0 + \sum_{k=1}^n B_{k}(t_i) \v_k 
\approx \boldsymbol{\psi}_0 + B_1(t_i)\v_1 + B_2(t_i)\v_2.
\end{equation}
The shape scores $B_k$ can be computed by a linear least-square fit. 
Fig.~\ref{fig_sperm}E displays the principal shape modes $\v_1$ and $\v_2$ as 
well as the superposition of a typical flagellar shape into these two modes. 
Using this procedure, the entire $n\times m$ data set $\Psi$ gets
projected onto an abstract shape space 
with just two axes representing the shape scores $B_1$ and $B_2$. 

\textit{Limit cycle reconstruction.}
Inspecting Fig.~\ref{fig_sperm}F, we find that the shape point cloud in shape 
space forms a closed loop: 
During each beat cycle, the shape points corresponding to subsequent 
flagellar shapes follow this shape circle to make one full turn. Thus, the 
shape space representation reflects the periodicity of the flagellar beat \cite{Geyer:2013}. 
As a next step, we can fit a closed curve to this point cloud, which defines 
a ``shape limit cycle''. 
We can then project each point of the cloud onto this limit cycle; this 
assignment is indicated as color-code in Fig.~\ref{fig_sperm}F. 
We parameterize the shape limit cycle by a phase angle $\varphi$ 
that advances by $2\pi$ after completing a full cycle. 
Furthermore, one can always assume that this phase angle increases uniformly 
along the curve \cite{Kralemann:2008}.
This procedure assigns a unique phase to each tracked flagellar shape and is 
equivalent to a binning of flagellar shapes according to shape similarity. 
We have thus arrived at a description of periodic flagellar beating in terms 
of a single phase variable that increases continuously
\begin{equation}
\label{eq_phi}
d\varphi /dt \approx \omega_0,
\end{equation}
where $\omega_0$ is the angular frequency of flagellar beating, see Fig.~\ref{fig_sperm}G.
Eq.~(\ref{eq_phi}) is a phase oscillator equation, 
which is a popular theoretical description for generic oscillators. 
This minimal description represents a starting point for more elaborate 
descriptions. 
For example, external forces have been shown to speed up or slow down the 
flagellar beat, which can be described by a single extra term in equation (\ref{eq_phi}) \cite{Geyer:2013}. 
Further, the scatter of the shape point cloud around the limit cycle of perfectly 
periodic flagellar beating reflects active fluctuations of the flagellar 
beat, which can be analyzed in a similar manner \cite{Ma:2014}. 

As an application of the shape-space representation, 
we follow \cite{Ma:2014} to define an instantaneous amplitude of the flagellar beat
as the radial distance $r(t)=[B_1(t)^2+B_2(t)^2]^{1/2}$
of a point in the $(B_1,B_2)$-shape space, 
normalized by the radial distance $\ol{r}$ of the corresponding point on the limit cycle of same phase.
We find that the fluctuations of this amplitude $r/\ol{r}$ are phase-dependent,
attaining minimal values during bend initiation at the proximal part of the flagellum, see Fig.~\ref{fig_sperm}H.
We argue that these amplitude fluctuations represent active fluctuations
stemming from the active motor dynamics inside the flagellum that drives flagellar waves.
As a test for the contribution from measurement noise, 
we added random perturbations to the tracking data, using known accuracies of tracking \cite{Riedel:2007}.
Phases and amplitudes computed for perturbed and unperturbed data were strongly correlated.

In conclusion, the reduction of the full data set $\Psi$ comprising $m$ 
feature dimensions to just a single phase variable involved a linear 
dimension reduction using principal component analysis to identify a shape 
limit cycle, followed by a problem-specific non-linear dimensionality 
reduction, the projection onto this limit cycle, 
to define phase and amplitude.
In future work, the shape space representation of the flagellar beat 
developed here can be used to quantify responses 
of the flagellar beat to mechanical or chemical stimuli.

\textit{Undulatory swimming with two shape modes.}
We will close this section by relating the results of our shape analysis to 
the hydrodynamics of flagellar swimming. 
For simplicity, we neglect variations of the flagellar beat and consider a 
perfect flagellar bending wave characterized by a 
``shape point'' circling along the ``shape limit cycle''. 
At the length scale of a sperm cell, inertia is negligible and the 
hydrodynamics of sperm swimming is governed by a low Reynolds number, which 
implies peculiar symmetries of the governing hydrodynamic equation (the 
Stokes equation) \cite{Lauga:2009}. 
In particular, the net displacement of the cell after one beat cycle will be 
independent of how fast the ``shape limit cycle'' is transversed. Further, 
playing the swimming stroke backwards in time would result exactly in a 
reversal of the motion. This implies that no net propulsion is possible for a 
reciprocal swimming stroke that looks alike when played forward or backward 
\cite{Purcell:1977}. The periodic modulation of just one shape mode is an 
example of such a reciprocal swimming stroke. In fact, the periodic 
modulation of one shape mode represents a standing wave, which does not allow 
for net propulsion, 
but implies that the sperm cells transverses a closed loop during a beat cycle, see Fig.~\ref{fig_sperm}J.
The superposition of two shape modes, however, represents a minimal system for swimming:
the superposition of two standing waves results in a traveling wave that breaks 
time-reversal symmetry and thus allows for net propulsion. The relation between 
standing and traveling waves can be illustrated by a minimal example of a 
trigonometric identity, which decomposes a traveling wave on the l.h.s. into 
two, periodically modulated shape modes of sinusoidal shape
\begin{equation}
\cos(\omega_0 t-2\pi s/\lambda) = \cos(\omega_0t)\cos(2\pi s/\lambda)+
\sin(\omega_0t)\sin(2\pi s/\lambda).
\end{equation}
In this minimal example, the shape modes would be given by sinusoidal 
standing wave profiles 
$\v_1(s)=\cos(2\pi s/\lambda)$ and $\v_2(s)=\sin(2\pi s/\lambda)$ with oscillating 
shape scores given by 
$B_1(t)=\cos(\omega_0 t)$ and $B_2(t)=\sin(\omega_0 t)$. 
Here, $\lambda$ corresponds to the wave-length of the waves. In the 
limit of small beat amplitudes, it can be formally shown that the net 
swimming speed of the cell is proportional to the area enclosed by the shape 
limit circle \cite{Shapere:1987}.

\subsection*{Shape and motility analysis of flatworms}

\begin{figure}
\begin{center}
\includegraphics[width=6in]{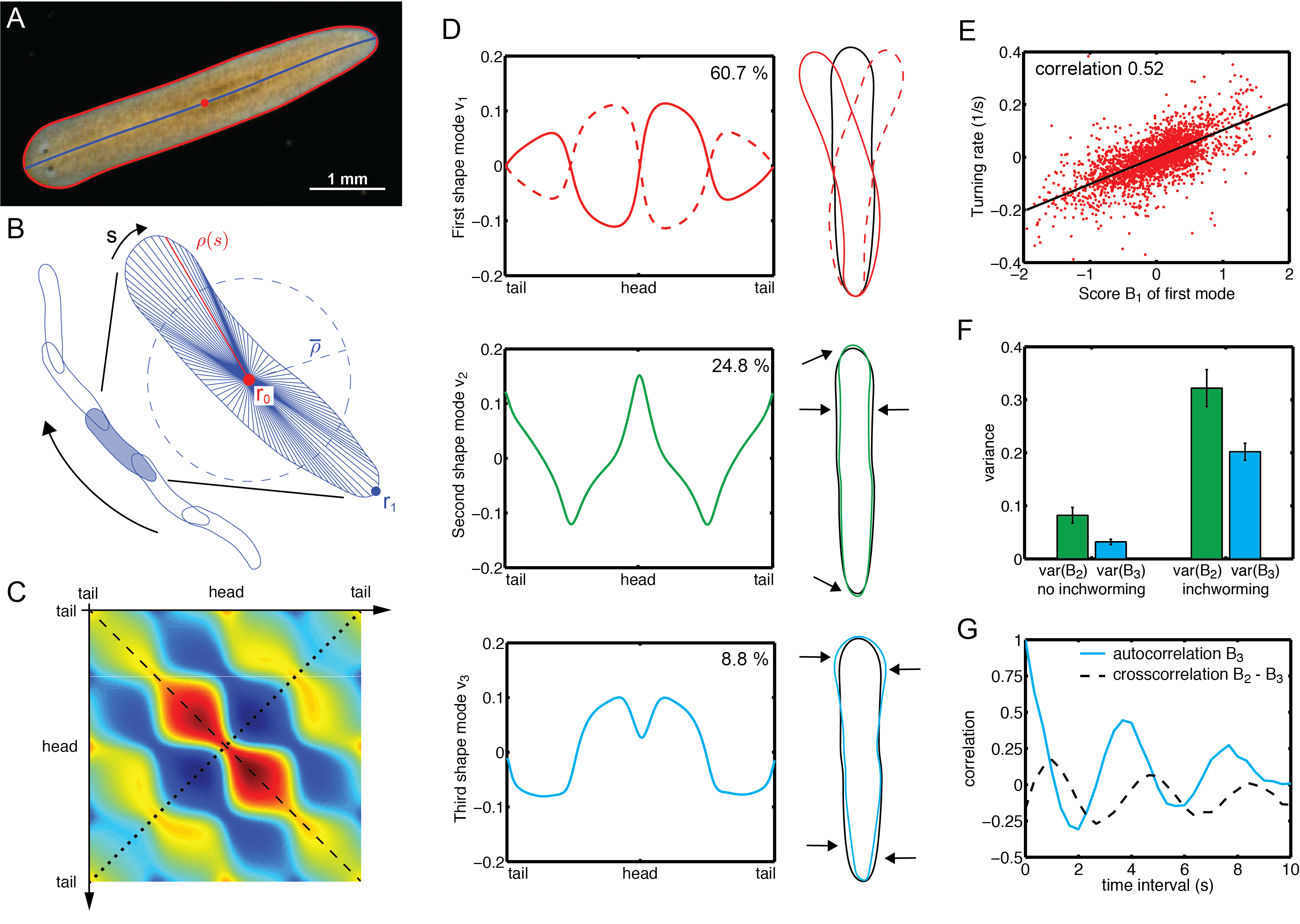}
\end{center}
\caption[]{
{\bf Three shape modes characterize projected flatworm body shape dynamics.}
\textbf{A.} 
Our custom-made MATLAB software tracks worms in movies and extracts worm 
boundary outline (red) and centerline (blue). 
\textbf{B.} 
The radial distance $\rho(s)$ between the boundary points and midpoint of the 
centerline ($r_0$, red dot) is calculated 
as a parameterization of worm shape.
We normalize the radial distance profile of each worm by the mean radius $\overline{\rho}$.
\textbf{C.}  The second symmetry axis (dotted line) of the covariance matrix corresponds to statistically symmetric behavior of the worm with respect to its midline.
\textbf{D.} 
The three shape modes with the largest eigenvalues account for 94\% of the 
shape variations. 
The first shape mode characterizes bending of the worm and alone accounts for 
61\% of the observed shape variance. 
On the top, we show its normalized radial profile on the left as well as the 
boundary outline corresponding to the superposition of the mean worm shape 
and this first shape mode (solid red: $B_1=1$, dashed red: $B_1=-1$, black: 
mean shape with $B_1=0$). The second shape mode describe lateral thinning ($B_2=0.3$), 
while the third shape mode corresponds unlike deformations of head and tail ($B_3=0.8$), 
giving the worm a wedge-shaped appearance. 
\textbf{E.} 
The first shape mode with score $B_1$ describing worm bending strongly 
correlates with the instantaneous turning rate of worm midpoint trajectories. 
\textbf{F.} We manually selected 30 movies where worms clearly show inch-worming and 50 movies with no inch-worming behavior. The variance of score $B_2$ and $B_3$ increases for the inch-worming worms. 
\textbf{G.} The autocorrelation of mode $B_3$ and the crosscorrelation between mode $B_2$ and mode $B_3$ reveals an inch-worming frequency of approximately $1/4\,\mathrm{Hz}$, hinting at generic behavioral patterns.
}
\label{fig_smed}
\end{figure}

We now apply shape mode analysis to time-lapse imaging data of
the flatworm \textit{Schmidtea mediterranea} (Fig.~\ref{fig_smed}A).
This animal is a popular model organism for 
studies on regeneration and growth \cite{Newmark:2002}. 
Flatworms (greek: \textit{Platyhelminthes}) represent some of the simplest organisms with bilateral body plan. 
Yet, they possess a distinct brain with two lobes, setting them apart from simpler worms like \textit{C. elegans}. 
Flatworms can steer their path in response to light, chemical stimuli, and temperature. 
Even a limited ability for learning has been proposed, including habituation and Pavlovian conditioning 
\cite{Shomrat:2013}. 
Hence, flatworms posses a sufficiently rich behavioral repertoire, whose control mechanisms are unknown to date. 
Flatworms are found in virtually all parts of the world, living in both salt- and freshwater, 
and include parasitic species like the cause of bilharzia. 
A subset of non-parasitic species, commonly referred to as `planarians', with
\textit{Schmidtea mediterranea} as a prominent member, 
is now entering the stage of modern model organisms 
to study regeneration, growth, and associated motility phenotypes \cite{Newmark:2002,Inoue:2004}.

Unlike the roundworm \textit{C. elegans}, planarians do not move by undulatory body motion. 
Instead, planarians glide over the substratum, 
being propelled by the beating of numerous short flagella 
(or cilia) that project from their multi-ciliated ventral epithelium. 
Planaria lack a rigid body wall and thus possess comparably soft bodies
that can deform significantly by muscle contractions. 
Thus, a continuous challenge in the field is the development of a reliable method 
to quantify shape variations of these soft-bodies animals. 
Below, we characterize their pronounced shape plasticity 
using shape mode analysis to characterize the outline of two-dimensional projections of their flat body. 
Similar characterization of outlines as closed curves are likely to be 
encountered in other contexts, 
such as the shape analysis of adherent or crawling cells \cite{Driscoll:2011}.

\textit{Worm handling and tracking.}
In the experiments, we use a clonal line of an asexual strain of \textit{Schmidtea mediterranea} 
\cite{SanchezAlvarado:2002, Benazzi:1972}. 
Worms were maintained at $20\degree$C as described in \cite{Cebria:2005} and
were starved for at least one week prior to imaging.
To monitor the 2D-projection of the worm body as in Fig.~\ref{fig_smed}A, 
we used a Nikon macroscope (AZ 100M, 0.5x objective) and a Nikon camera set-up (DS-Fi1, frame rate 3 Hz, total observation period 15 s, resolution 1280 x 960 pixel). 
The flatworms were placed one at a time into a plastic petri dish ($90$ mm), clean petri dishes were used for each experimental series (comprising $2-3$ movies of $20-30$ worms). 
After being exposed to light, worms displayed a typical flight response. 
Movies were analyzed off-line using custom-made MATLAB software.
A first shape proxy was determined from background-corrected movie frames
via edge detection using a canny-filter, followed by a dilation-erosion cycle.
In a subsequent refinement step, the worm perimeter was adjusted by
finding the steepest drop in intensity along directions transverse to the perimeter proxy.
As a result we were able to automatically track the boundary outline (red) as well as the centerline (blue) of worms with sub-pixel accuracy in a very robust manner, see Fig.~\ref{fig_smed}A.

\textit{Radial profiles characterize non-convex outlines.}
In analyzing the tracked outline shapes, we face the challenge of characterizing the shape of 
closed, planar curves. 
For non-convex shapes, this can be non-trivial. 
We describe a closed curve by a position vector $\r(s)$ as a function of 
arc-length $s$ along its circumference, see Fig.~\ref{fig_smed}B. 
We use the tip of the worm tail as a distinguished reference point $\r_1$ 
that specifies the position of $s=0$. 
We further specify a center point $\r_0$, using the midpoint of the tracked 
centerline of the worms. 
The profile of radial distances $\rho(s)=|\r(s)-\r_0|$ measured with 
respect to the center point $\r_0$ characterizes outline shape, even for non-convex outlines. 
 Shapes of convex curves might also be characterized by a profile of radial 
distances $\rho(\varphi)$ as a function of a polar angle $\varphi$. 
However, this definition does not generalize to non-convex curves 
(or, more precisely, to curves that are not radially convex with respect to $\r_0$). 
To adjust for different worm sizes, we normalize the radial distance profiles 
by the mean radius 
$\overline{\rho}=\langle \rho(s)\rangle$ as 
$\hat{\rho}=\rho(s)/\overline{\rho}$ and plot it as a function of normalized 
arc-length $\hat{s}=s/L$, where $L$ is the total length of the circumference.
As a mathematical side-note, we remark that using the signed curvature 
$\kappa(s)=(d^2\r(s)/ds^2)\cdot(d\r(s)/ds)$ along the circumference, instead 
of the radial distance profile $\rho(s)$, would amount to a significant 
disadvantage: The property that a certain curvature profile actually 
corresponds to a closed curve imposes a non-trivial constraint on the set of 
admissible curvature profiles. For the normalized radial distance profiles, 
however, there is a continuous range of distance profiles that correspond to 
closed curves, making this choice of definition more suitable for applying 
linear decomposition techniques such as shape mode analysis. In fact, given a 
particular normalized radial distance profile, the corresponding 
circumference length $L/\overline{\rho}$ is reconstructed self-consistently 
by the requirement that the associated curve must close on itself.

\textit{A bending mode and two width-changing modes.}
We extracted $n=29\;993$ worm outlines from a total of 745 analyzed movies.
We computed normalized radial distance profiles as described above, each 
profile being represented by $m=200$ radii, resulting in a large $n\times m$ data matrix.
From the average of all radial profiles, 
we define a mean worm shape that averages out shape variations, 
see Fig.~\ref{fig_smed}D (right inset,black). 
Next, we computed the covariance matrix $\C$ between the individual radial profiles, 
using the centered (mean-corrected) data matrix,
Fig.~\ref{fig_smed}C. 
The symmetry of the covariance matrix along the dotted diagonal 
shows that shape variations are statistically symmetric with respect to the worm midline.
Again, the eigenvectors corresponding to the largest eigenvalues of this 
matrix are those with maximal descriptive power for shape variance. 
Fig.~\ref{fig_smed}D shows the first three shape modes, 
which together account for $94\%$ of the observed shape variance.
We find that the dominant shape mode $\v_1$ is anti-symmetric,
describing an overall bending of the worm. 
In contrast, the second and third mode describe symmetric width changes of the worm:
The second shape mode $\v_2$ characterizes a lateral thinning 
of the worm associated with a pointy head and tail. 
Correspondingly, a negative contribution of the second shape mode with $B_2<0$ describes lateral 
thickening of the worm (with slightly more roundish head and tail).
The third shape mode $\v_3$, finally, is also symmetric and is associated
with unlike deformations of head and tail, giving the worm a wedge-like appearance. 
Superpositions of these three shape modes describe in-plane bending of the worms, 
and a complex width dynamics of head and tail.

\textit{The bending mode characterizes turning.}
Next, we investigated the relationship between flatworm shape and motility. 
Flatworms employ numerous beating cilia on their ventral epithelium to glide on surfaces.
We observe that worms actively regulate their gliding speed over a 
considerable range of $0.3-1.7\,\mathrm{mm/s}$,
in quantitative agreement with earlier work \cite{Talbot:2011}.
Yet, we did not observe pronounced correlations between shape dynamics and gliding speed (not shown).
This is consistent with the notion that muscle contractions play a minor role in the generation of normal gliding motility.
However, we find that shape changes control the direction of gliding motility and thus steer the worm's path:
Fig.~\ref{fig_smed}E displays a significant correlation between the rate of turning along the worm trajectory 
and the first shape score $B_1$, which characterizes bending of the worm.
The sign and magnitude of this ``bending score'' directly relates  to the direction and rate of turning. 
For simplicity,
we had restricted the analysis to a medium size range of $8$-$10\,\mathrm{mm}$ length, 
analogous results are found for other size ranges.

\textit{The second and third modes characterize inch-worming.}
In addition to cilia-driven gliding motility, 
flatworms employ a second, cilia-independent motility pattern known as inch-worming, 
which provides a back-up motility system in case of dysfunctional cilia \cite{Rink:2009}
or as an escape response. 
To test whether modes two and three might relate to this second motility pattern, 
we analyzed movies of small worms known to engage more frequently in this kind of behaviour. 

We therefore manually classified $80$ movies
of worms smaller than $0.9$ mm that have been starved for 10 weeks, 
yielding a number $30$ inch-worming and $50$ non-inch-worming worms
for a differentiated motility analysis (cases of ambiguity were not included).
We find that the second and third shape mode, 
which characterize dynamic variations in body width,
are indeed more pronounced in inch-worming worms, see Fig.~\ref{fig_smed}F.
Next, we computed the temporal autocorrelation of time series of the
second shape mode $B_3$, see Fig.~\ref{fig_smed}G (solid blue).
We observe stereotypical shape oscillations with a characteristic frequency
of 0.26 Hz. 
From the cross-correlation between $B_3$ and $B_2$ in Fig.~\ref{fig_smed}G (dashed black), 
we find that both shape scores oscillate with a common frequency and relative phase lag of $-0.6\,\pi$ (where $B_2$ lags behind). 
Thus, both shape modes act together in an orchestrated manner to faciliate inch-worming,
hinting at coordinated muscle movements and periodic neuronal activity patterns.

In conclusion, 
we identified different shape modes 
that can characterize different fundamental types of motility in a quantitative manner. 
The characteristic periodic shape dynamics associated with inch-worming
posses the question about underlying generic patterns of neuronal and muscular activity.

\begin{figure}[!ht]
\begin{center}
\includegraphics[width=6in]{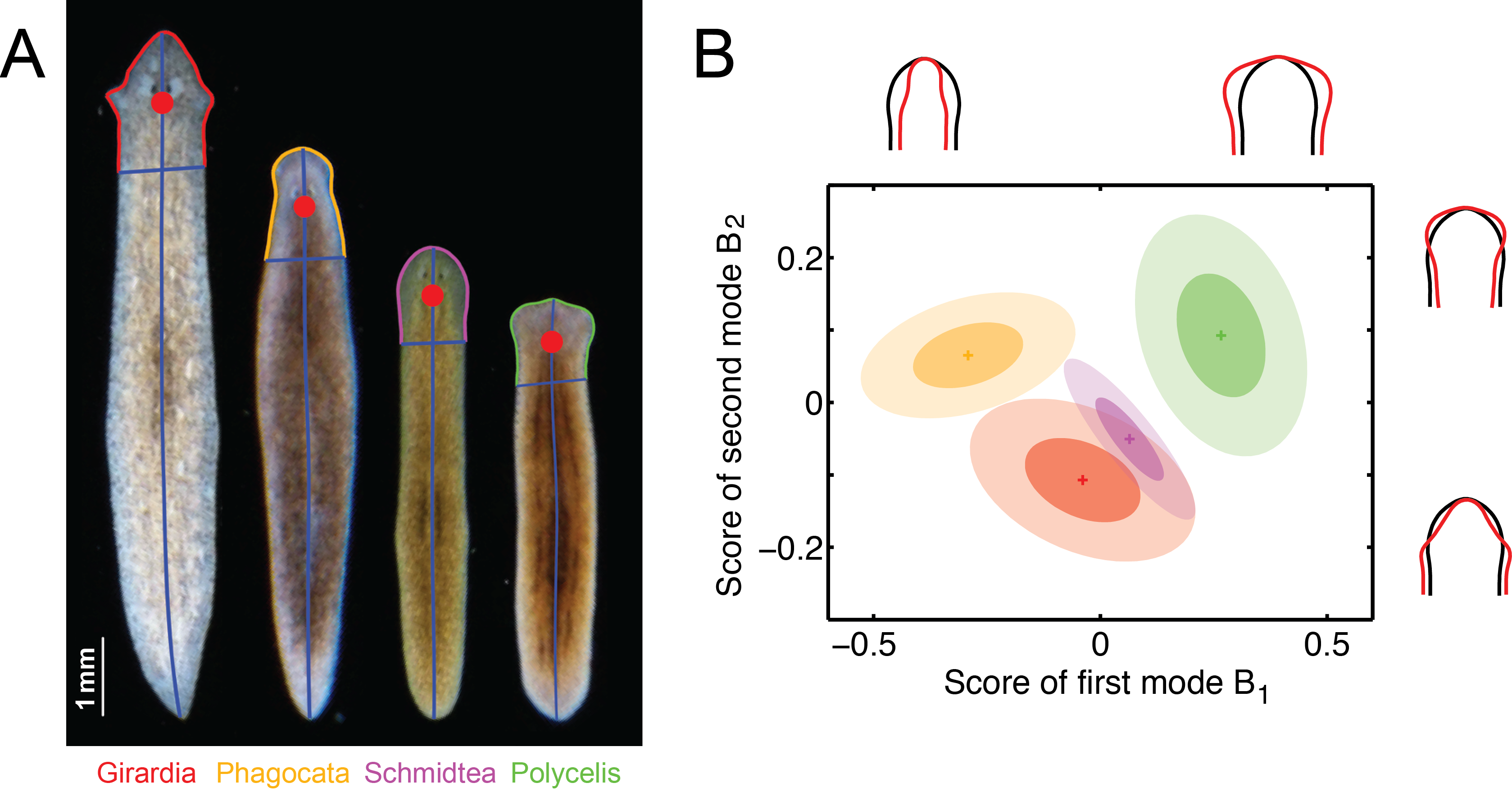}
\end{center}
\caption[]{
{\bf Distinguishing head morphologies of four different flatworm species.}
\textbf{A.} Application of our method to parametrize head morphology of four different flatworm species. For each species, time-lapse sequences of $4$ different worms were recorded as two independent runs of duration $16$ frames. The head is defined as most anterior $20\%$ of the worm body. Radial distances $\rho(s)$ are computed with respect to the midpoint of the head (red dot at $10\%$ of the worm length from the tip of the head).
\textbf{B.} By applying PCA to this multi-species data set, we obtain two shape modes, which together account for $88\%$ of the shape variability. Deformations of the mean shape with respect to the the two modes are shown (black: mean shape, red: superposition of mean shape and first mode with $B_1=\pm0.4$ and second mode with $B_2=\pm0.2$, respectively). We represent head morphology of the four species in a combined shape space of these two modes. Average head shapes for each species are indicated by crosses, with ellipses of variance including $68\%$ (dark color) and $95\%$ (light color) of motility-associated shape variability, respectively.}
\label{fig_species}
\end{figure}

\textit{PCA discriminates flatworm species.}
Having developed tools to measure shape changes of the same animal over time, 
we next explored the utility of shape mode analysis in shape comparisons between different animals.
The model species \textit{Schmidtea mediterranea} is but one of many hundred flatworm species existing worldwide \cite{Liu:2013}. 
The taxonomic identification of planarian species is challenging, relying largely on the time-consuming mapping of internal characters.
The availability of quantitative bodyplan morphological parameters would be of interest in this context. 
Having available a large live collection of planarian species, we choose four species 
representing the genera \textit{Girardia}, \textit{Phagocata}, \textit{Schmidtea} and \textit{Polycelis}.
Besides potentially size-dependent variations in aspect ratio, 
the four species differ by their characteristic head shapes, see Fig.~\ref{fig_species}A. 
Accordingly, we restricted  shape analysis to the head region only 
(defined as the most anterior $20\%$ of the worm body).
We characterized each head shape by a vector of distances 
from the midpoint of the head (red dot, $10\%$ of the worm length from the tip of the head)
to the outline $\rho(s)$ of the head region and proceeded as above. 
We found that the first two eigenmodes captured $88\%$ of head shape variability 
within this multi-species data set. 
Fig.~\ref{fig_species}B shows species-specific mean shapes for each of the four species in a combined head shape space, as well as ellipses of variance covering $68\%$ (dark color) and $95\%$ (light color) of motility-associated shape variability, respectively. 
This comparison of flatworm species representing four genera
illustrates linear dimensionality reduction as a simple means to map morphological differences across species.

\subsection*{Discussion}

Using two biological examples,
swimming sperm and gliding flatworms,
we demonstrated shape mode analysis as a 
versatile tool to characterize morphological shapes and its dynamical changes. 
In both cases, we obtained a low-dimensional description of organism shape.
Our observation that complex shapes dynamics can be concisely described by just a few shape scores
corroborates the high coordination of molecular motor activity in the sperm flagellum, 
as well as contraction of muscles in flatworms during both gliding and inch-worming motility, respectively.

In the case of a beating flagellum, shape mode analysis revealed a limit cycle 
that characterizes the periodicity of the beat.
This limit cycle allowed the definition of a flagellar phase that rectifies the 
progression through a periodic sequence of shapes
as well as the quantification of noisy deviations from perfectly periodic shape dynamics. 
In the appendix, we comment on the challenges to deal with the periodicity of angular data.

In the case of flatworms, 
shape mode analysis concisely characterizes a behavioral repertoire and the associated body shape dynamics. 
It is known that flatworms employ two distinct motility mechanisms: 
(i) gliding motility, relying on beating of their ventral cilia with occasional turns, and,
(ii) inch-worming, which is driven by muscle contractions \cite{Rink:2009, Rompolas:2010}.
We find that bending and turning maneuvers are strongly correlated, 
revealing a generic mechanism for steering.
Furthermore, we quantitatively analyzed the motility mechanisms of inch-worming,
which is evoked in case of dysfunctional cilia \cite{Rink:2009, Rompolas:2010} or as an escape response.
We observe a concerted shape dynamics of lateral thinning of head and tail 
with a characteristic period of about $4\,\mathrm{s}$.
Our analysis can serve as basis for future studies of generic behavioral responses in planarians 
and underlying patterns of neuronal and muscular activity,
irrespective of their higher level of complexity \cite{Stephens:2008,Stephens:2010}
compared to other model organisms such as \textit{C. elegans}. 
Previous studies of planarian motility had focused 
on coarse-grained motility parameters such as net speed or the 
mean-squared-displacement of worm tracks \cite{Talbot:2011}. 
To the best of our knowledge, 
our study represents the first application of shape mode analysis to flatworm motility,
linking shape and motion in a quantitative manner, 
thus enabling the characterization of motility phenotypes.

Additionally, our method presents a simple means to compare and distinguish different flatworm species.
Further refinements of the method that take into account the whole body shape 
could generate a useful supplement of taxonomic traits to help in the classification of new planarian species. 
Further, the ability to precisely quantify differences in head shape 
now enable the dissection of the underlying molecular pathways that control morphogenesis. 
Based on the principal components defining head shape, 
it is conceivable that planarian head morphogenesis is mainly controlled by two molecular networks: 
One controlling maximal head width at the position of the auricles 
and a second one determining the posterior displacement of the point of maximal head width. 
The availability of transcriptome sequence information for these species (Liu et al, in preparation) 
will now allow testing of this hypothesis, \textit{e.g.}, 
by RNAi screens with shape mode analysis as read-out 
or systematic expression level comparisons in head transcriptomes of species with different head morphologies. 
Similar inter-species comparisons of beak morphology in Darwin finches 
could be correlated with the ecological niche of the animals \cite{Shoval:2012}. 
The corners of the observed shape set corresponded 
to archetypical species that are highly specialized to a narrow environmental niche,
while species corresponding to interior points of the shape set represent generalists, 
whose fitness is optimized simultaneously for several traits. 
It will be interesting to test similar hypotheses for flatworm species, 
some of which inhabit extreme environments.

\subsection*{Mathematical appendix:\\ 
PCA for angular data and kernel methods}

\begin{figure}[!ht]
\begin{center}
\includegraphics[width=6in]{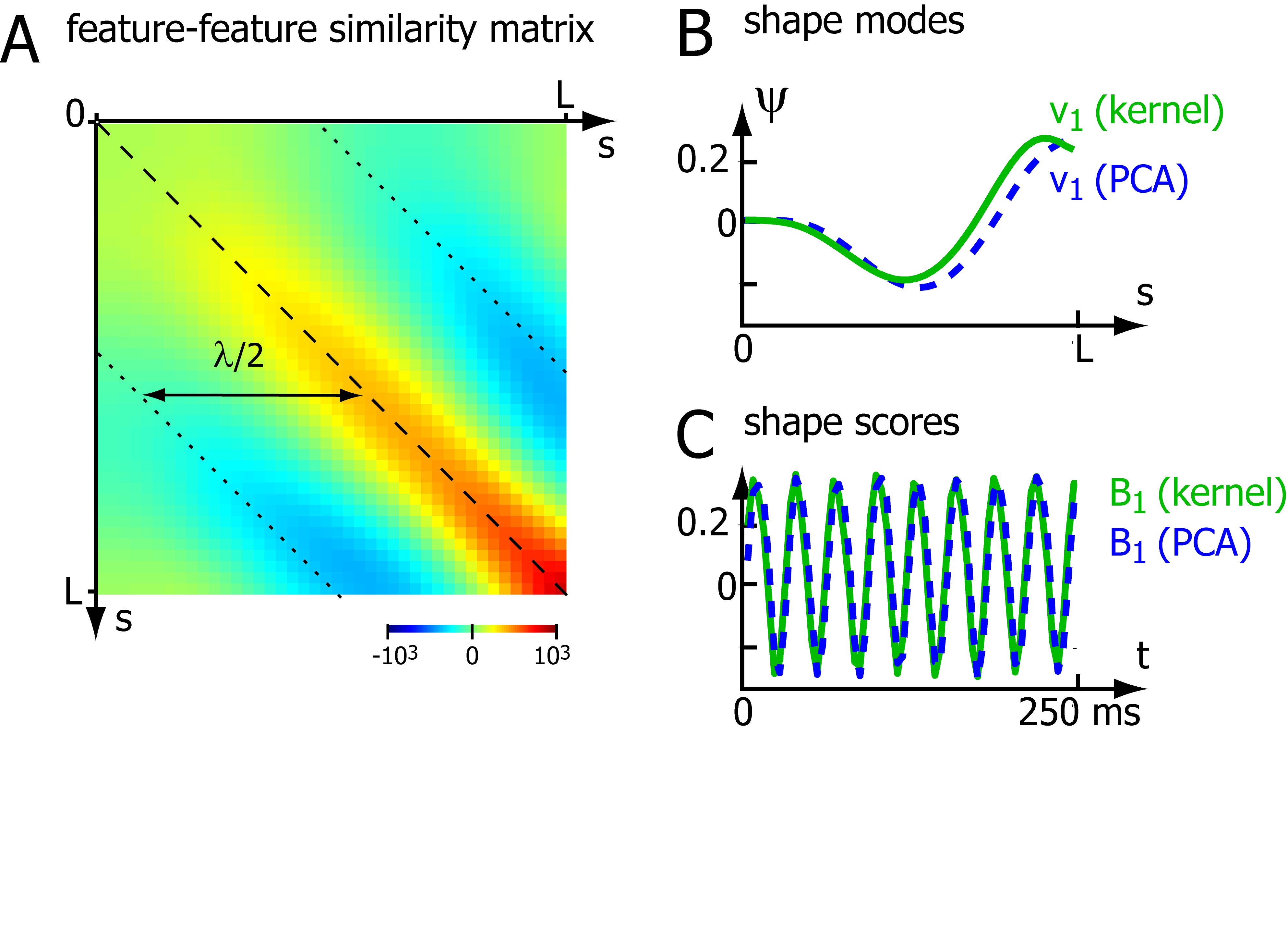}
\end{center}
\caption[]{
\label{fig_kernel_PCA}
{\bf Principal component analysis for angular data using kernel PCA.}
\textbf{A.} 
Centered feature-feature similarity matrix $\C^0$ according to eq.~(\ref{eq_kernel}) for the sperm tangent angle data.
\textbf{B.} 
First shape mode for the kernel method (green) compared to the first shape mode as obtained by linear PCA (blue dashed).
\textbf{C.} 
Corresponding shape scores $B_1(t)$ as a function of measurement time
for both the kernel method (green) and for linear PCA (blue dashed).
}
\end{figure}

We discuss an extension of principal component analysis 
using a distance kernel, which is particularly suited for the analysis of angular data.
Linear operations on angular data can be problematic,
\textit{e.g.} if angles jump by $2\pi$.
We define a $n\times n$ feature-feature similarity matrix that accounts for the
$2\pi$-ambiguity of angle data
\begin{equation}
\label{eq_kernel}
C_{ij}=\sum_k \cos[ \psi(s_i,t_k)-\psi(s_j,t_k) ].
\end{equation}
Rows and columns of this matrix do not automatically average to zero, 
so kernel centering \cite{Scholkopf:kernel} has to be applied,
$$
C^0_{ij}=C_{ij}-\frac{1}{n}\sum_{l=1}^n C_{il}-\frac{1}{n}\sum_{k=1}^n C_{kj} +\frac{1}{n^2}\sum_{k,l=1}^n C_{kl}.
$$
We can now proceed as below eq.~(\ref{eq_C}),
obtaining shape modes $\v_k$ from the eigenvectors of the matrix $\C^0$.
(Without kernel centering, the resultant shape modes would comprise 
a contribution from the non-zero average of all the measurements \cite{Cannistraci:2013}.)
Shape scores $B_k(t_j)$ can be defined by maximizing the 
similarity measure
$$
\sum_{i=1}^n \cos[\psi(s_i,t_j)-\psi_0(s_i)-B_1(t_j)\v_1(s_i)-B_2(t_j)\v_2(s_i)].
$$
Here, the `mean flagellar shape' $\psi_0(s_i)$ is defined using the 
circular mean $\psi_0(s_l)=\mathrm{arg} \sum_{j=1}^m \exp[i\psi(s_l,t_j)]$.
Fig. \ref{fig_kernel_PCA} compares the first shape mode and its scores
for this kernel PCA and linear PCA as considered in the main text.

If one is only interested in shape scores, but not the corresponding shape modes,
an alternative approach would be to use a $m\times m$ measurement-measurement similarity matrix 
$$
G_{ij}=\sum_{l=1}^n \cos[ \psi(s_l,t_i)-\psi(s_l,t_j) ],
$$
known as a Gram's matrix \cite{Scholkopf:kernel}.
The eigenvalues $d_1>d_2>\ldots$ and
eigenvectors $u_1$, $u_2$, $\ldots$ of the kernel-centered Gram matrix $\G^0$
provide a proxy for the shape scores via
$B_k(t_j)=\sqrt{d_k} u_{k,j}$.

\begin{figure}[!ht]
 \begin{center}
\includegraphics[width=5in]{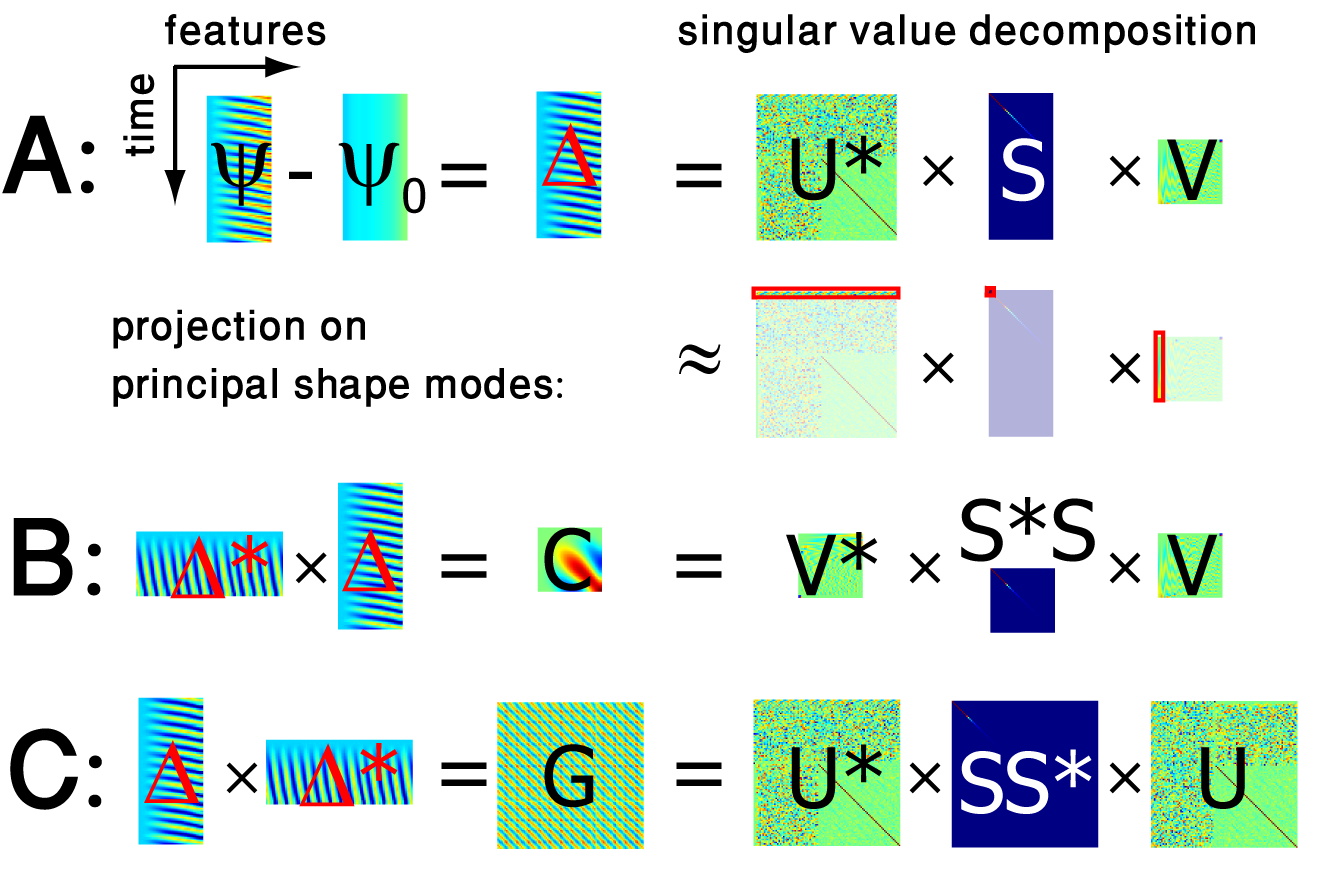}
 \end{center}
 \caption[]{
 {\bf The mathematics behind principal component analysis (PCA).}
 \textbf{A.} 
 For illustration, we start with a $n\times m$ measurement matrix $\psi$ 
 featuring the beat of a sperm flagellum with $n=100$ measurement (rows) and 
 tangent angles at $m=41$ equidistant positions along the flagellar centerline 
 (columns). Subtracting the mean defines the centered $n\times m$ data 
 matrix $\Delta$. The mathematical technique of singular value decomposition 
 factors the data matrix $\Delta$ into a product of a unitary $n\times n$ 
 matrix $\U^\ast$, a ``diagonal'' $n\times m$ matrix $S$ that has non-zero 
 entries only along its diagonal, and a unitary $m\times m$ matrix $V$. 
 Singular value decomposition may be regarded as a generalization of the usual 
 eigensystem decomposition of symmetric square matrices to non-square 
 matrices. A unitary $n\times n$ matrix $\U$ generalizes the concept of a 
 rotation matrix to $n$-dimensional space.
; it is defined by $\U^\ast\U=\U\U^\ast$ being equal to the identity matrix. 
 \textit{Second row:} A restriction to the top-$k$ singular values
 defines sub-matrices of $\U^\ast$, $\S$, $\V$ of dimensions
 $n\times k$, $k\times k$, $k\times m$, respectively, 
 whose product represents a useful approximation of the full factorization
 that reduces $m$ feature dimensions to only $k$ shape modes.
 \textbf{B.} 
 The $m\times m$ feature-feature covariance matrix $\C$ is defined in terms 
 of the centered data matrix $\Delta$. It can be written as a product of 
 a diagonal matrix $\D=S^\ast S$, whose diagonal features the eigenvalues of 
 $\C$ and a unitary $m\times m$-matrix $\V$ whose columns correspond to the 
 respective (left) eigenvectors of $\C$. This matrix $\V$ is exactly the same 
 as previously encountered in the singular value decomposition of $\Delta$. 
 \textbf{C.} 
 Similarly, the $m\times m$ measurement-measurement covariance matrix 
 $\G=\Delta\Delta^\ast$, known as the Gram matrix, can be decomposed using a 
 diagonal $m\times m$ matrix $\S\S^\ast$ and a unitary matrix $\U$. 
 Importantly, the rows of $\V$ comprise just the $m$ shape modes of the data 
 matrix $\Delta$ as defined by linear PCA, while the columns of the matrix 
 $\B=\U^\ast\S$ yield the corresponding shape scores. }
 \label{fig_mat}
 \end{figure}

A mathematical motivation for the use of such kernel methods
stems from the fact eq.~(\ref{eq_C}) can be interpreted as a special case of a similarity kernel.
For linear PCA, the eigenvector decomposition of 
the feature-feature covariance matrix $\C=\Delta^\ast\Delta$ 
and that of the measurement-measurement covariance matrix $\G=\Delta\Delta^\ast$ 
yield analogous results as can be shown using the 
singular value decomposition of the mean centered data matrix $\Delta=\Psi-\Psi_0$
\begin{equation}
\label{eq_SingValDec}
\Delta=\U^\ast \S \V.
\end{equation}
Here, $\U$ and $\V$ are unitary $n\times n$ and $m\times m$ matrices, respectively, 
and $\S$ is a diagonal $n\times m$ matrix.
From eq.~(\ref{eq_SingValDec}), we readily find 
$\C=\V^\ast\; \D_C\; \V$
and
$\G=\U^\ast\; \D_G\; \U$,
where $D_C=\S^\ast\S$ and $D_G=\S\S^\ast$
are diagonal matrices with the same eigenvalues.
These matrix decompositions are illustrated in Fig. \ref{fig_mat} for linear PCA on sperm data.

Using a nonlinear similarity measure as in eq.~(\ref{eq_kernel}) breaks the exact correspondence
between PCA and kernel PCA based on the measurement-measurement covariance matrix. 
Nevertheless, the use of kernels allows to analyze more complicated data sets
and depicts the road to nonlinear dimension reduction methods \cite{Scholkopf:kernel}.
In fact, several nonlinear dimensionality reduction algorithms rely 
on kernel PCA, including the popular Isomap algorithm \cite{Tenenbaum:2000}.
Such algorithms have been used for automated frame-sorting, 
including flagellar video-microscopy \cite{Bayly:2011}.
Additionally, the concept of a Gram matrix is used in multi-dimensional-scaling
to reconstruct embeddings into a high-dimensional feature space using 
only a Gram matrix of mutual distances between individual measurements.

\section*{Supporting Online Material}
A Matlab script is a available for download that illustrates the method of 
shape mode analysis by principal component analysis, 
and the reconstruction of a limit cycle as shown in Fig.~\ref{fig:sperm}.

\section*{Acknowledgments}

We thank Johannes Baumgart and Carlos V. Cannistraci 
for stimulating discussions and Frank J{\"u}licher for continuous support. 
We are grateful to M.~Vila-Farr\'e for providing flatworm species and N.~Alt for carefully imaging the flatworms.

\bibliography{ShapeModesPaper}

\clearpage

\section*{Figure Legends}

\clearpage

\begin{figure}[!ht]
\begin{center}
\includegraphics[width=6in]{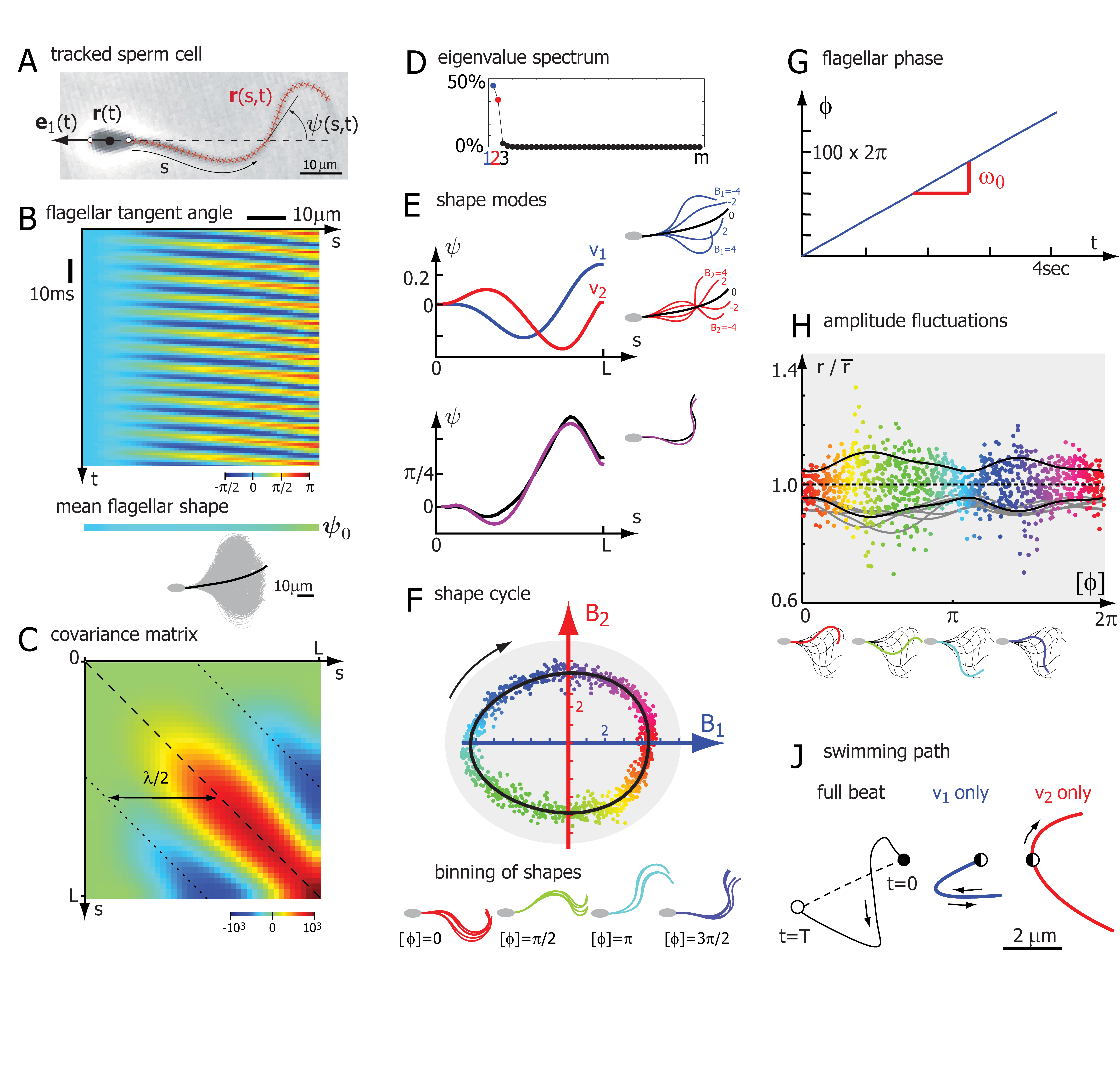}
\end{center}
\caption[]{
\label{fig:sperm}
{\bf Principal shape modes of sperm flagellar beating.} 
(continued on next page).
}
\label{fig_sperm}
\end{figure}

\clearpage

\hrule
\vspace{2mm}
\setlength{\parindent}{0mm}
{\bf Figure \ref{fig_sperm}:} 
{\bf Principal shape modes of sperm flagellar beating}
(continued from previous page).\\
\textbf{A.} 
High-precision tracking of planar flagellar centerline shapes ($\mathbf{r}(s,t)$, red) 
are characterized by their tangent angle $\psi(s,t)$ as a function of 
arc-length $s$ along the flagellum.
\textbf{B.} The time-evolution of this flagellar tangent angle is shown as a 
kymograph. 
The periodicity of the flagellar beat is reflected by the regular stripe 
patterns in this kymograph; the slope of these stripes is related to the 
propagation of bending waves along the flagellum from base to tip. By 
averaging over the time-dimension, we define a mean flagellar shape 
characterized by a tangent angle profile $\boldsymbol{\psi_0}$. For illustration, this 
mean flagellar shape is shown in black superimposed to $n=1024$ tracked 
flagellar shapes (grey).
\textbf{C.} We define a feature-feature covariance matrix $\C$ from the 
centered tangent angle data matrix as explained in the text. The 
negative correlation at arc-length distance $\lambda/2$ reflects the 
half-wavelength of the flagellar bending waves.
\textbf{D.} The normalized eigenvalue 
spectrum of the covariance matrix $\mathrm{C}$ sharply drops after the second 
eigenvalue, implying that the eigenvectors corresponding to the first two 
eigenvalues together account for 97\% of the observed variance in the tangent 
angle data.
\textbf{E.} Using principal component analysis, we define two principal shape 
modes (blue, red), which correspond precisely to the two maximal eigenvalues 
of the covariance matrix $\mathrm{C}$ in panel C. The lower plot shows the 
reconstruction of a tracked flagellar shape (black) by a superposition of the 
mean flagellar shape and these two principal shape modes (magenta). In 
addition to tangent angle profiles, respective flagellar shapes are shown on 
the right.
\textbf{F.} Each tracked flagellar shape can now be assigned a pair of shape 
scores $B_1$ and $B_2$, indicating the relative weight of the two 
principal shape modes in reconstruction this shape. This defines a 
two-dimensional abstract shape space. A sequence of shapes corresponds to a 
point cloud in this shape space. We find that these point form a closed loop, 
reflecting the periodicity of the flagellar beat. We can define a shape limit 
cycle by fitting a curve to the point cloud. By projecting the shape points 
on this shape limit cycle, we can assign a unique flagellar phase $[\varphi]$ 
modulo $2\pi$ to each shape. This procedure amounts to a binning of flagellar 
shapes according to shape similarity.
\textbf{G.} By requiring that the phase variable $\varphi$ should change 
continuously, we obtain a representation of the beating flagellum as a phase 
oscillator. The flagellar phase increases at a rate equal to the frequency of 
the flagellar beat and rectifies the progression through subsequent beat 
cycles by increasing by $2\pi$.
\textbf{H.} 
Amplitude fluctuations of flagellar beating as a function of flagellar phase.
An instantaneous amplitude of the flagellar beat is defined
as the radial distance $r(t)$ of a point in the $(B_1,B_2)$-shape space, 
normalized by the radial distance $\ol{r}(\varphi(t))$ of the corresponding point on the limit cycle of same phase.
A phase-dependent standard deviation was fitted to the data (black solid line).
Also shown are fits for $6$ additional cells
(gray; the position of $\varphi=0$ was defined using a common set of shape modes).
\textbf{J.}
Swimming path of the head center during one beat cycle computed 
for the flagellar wave given by the shape limit cycle (panel F)
using resistive force theory \cite{Gray:1955b} as described previously \cite{Friedrich:2010}.
The path is characterized by a wiggling motion of the head superimposed to net propulsion.
For a `standing wave' beat pattern characterized by the oscillation of only one shape mode,
net propulsion vanishes.

\clearpage
\begin{table}[!p]
  \centering
  \begin{tabular}{l|l|l|l}
  $V_i$ & $B_i$ & $D_i$ & Ref.\\
  \hline
   coefficients, loadings & principal components & eigenvalues &  \cite{Jolliffe:2005}\\
   characteristic vectors, eigenvectors & z-scores & characteristic roots, latent roots & \cite{Jackson:2005}\\
   eigenvectors & amplitudes & eigenvalues &\cite{Stephens:2008}\\
   coefficients, loadings & scores & latents, eigenvalues & 
   MATLAB \cite{MATLAB:2013} \\
  \end{tabular}
  \caption{Principal component analysis is used across different disciplines, giving rise to a diverse terminology, which is summarized here.}
  \label{tab:terminology}
\end{table}

\end{document}